\documentclass{article}
\usepackage{spconf,amsmath,graphicx}
\usepackage{amssymb}
\usepackage{xcolor}
\usepackage{algorithm}
\usepackage{algpseudocode}
\usepackage{hyperref}
\usepackage{diagbox}
\usepackage{multicol}
\usepackage{multirow}

\usepackage{tikz}
\usepackage{pgfplots}
\pgfplotsset{compat=1.9} 
\usetikzlibrary{arrows.meta}
\usetikzlibrary{calc}
\usetikzlibrary{positioning}

\definecolor{pltblue}{RGB}{31, 119, 180}

\title{On the Possible Detectability of Image-in-Image Steganography}

\twoauthors
  {Antoine Mallet}
	{Univ. Lille, CNRS\\
   UMR 9189 CRIStAL\\
	 F-59000 Lille, France}
  {Patrick Bas}
	{Univ. Lille, CNRS\\
   UMR 9189 CRIStAL\\
   F-59000 Lille, France}
\begin{document}
%
  \maketitle
  \begin{abstract}

    This paper investigates the detectability of popular image-in-image steganography schemes~\cite{jing2021hinet,yang2024pris,guan2022deepmih,baluja2019hiding,weng2019high}.
    In this paradigm, the payload is usually an image of the same size as the Cover image, leading to very high embedding rates.
    We first show that the embedding yields a mixing process that is easily identifiable by independent component analysis.
    We then propose a simple, interpretable steganalysis method based on the first four moments of the independent components estimated from the wavelet decomposition of the images, which are used to distinguish between the distributions of Cover and Stego components.
    Experimental results demonstrate the efficiency of the proposed method, with eight-dimensional input vectors attaining up to $84.6\%$ accuracy. This vulnerability analysis is supported by two other facts: the use of keyless extraction networks and the high detectability w.r.t. classical steganalysis methods, such as the SRM combined with support vector machines, which attains over $99\%$ accuracy.

  \end{abstract}
%


\section{Introduction}
\label{sec:introduction}

  \begin{figure}[t!]
    \centering
    \includegraphics[width=\columnwidth]{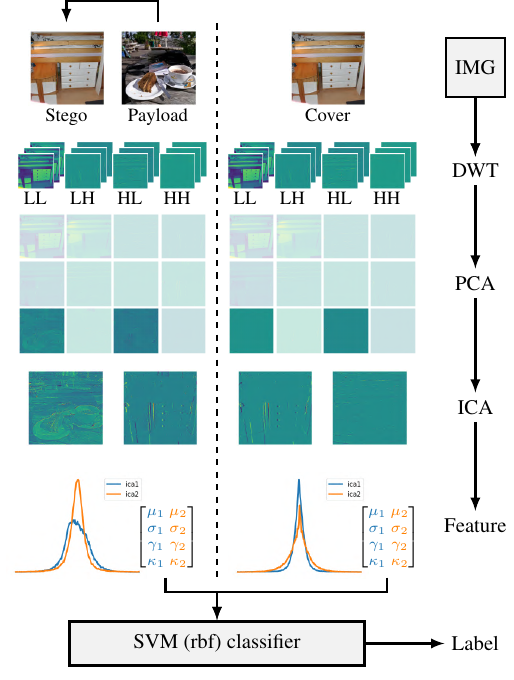}
    \caption{Overview of the proposed steganalysis method. First, a discrete wavelet transform (DWT) is applied to the input image.
    Then, principal component analysis (PCA) is performed to extract weak components.
    Independent component analysis (ICA) is applied to extract independent sources that are semantically close to the Cover and Payload images. 
    Finally, the first four moments (mean $\mu$, standard deviation $\sigma$, skewness $\gamma$, and kurtosis $\kappa$) of the estimated independent components are computed to form a discriminative yet interpretable feature vector, which is used to train an SVM classifier for detection.
    }
    \label{fig:method_overview}
    \vspace{-0.5cm}
  \end{figure}

  Steganography is the science of hiding a secret message within an innocuous-looking medium.
  To be effective, a steganographic scheme must ensure that the modifications brought to the Cover medium are as imperceptible as possible 
  to the scrutiny of a potential adversary, called the steganalyzer.
  One of the main challenges of steganography, in this pursuit, is to design schemes that minimize the statistical detectability of the Stego object, 
  while maximizing the size of the embedded information relative to the size of the Cover object, which we call the Payload.

  In the last few years, a branch of steganographic schemes has aimed at hiding an image within another image. 
  In this setup, the Cover and the secret message are both images.
  Contrary to classical steganography schemes, where the secret message is a bitstream much smaller than the Cover object, 
  image-in-image steganography schemes aim at embedding a payload of comparable size to the Cover image. 
  To do so, these schemes usually rely on deep learning architectures, such as encoder-decoder architectures~\cite{zhu2018hidden, 
  wu2018stegnet, baluja2019hiding}, invertible neural networks (INN)~\cite{jing2021hinet, lu2021large, xu2022robust, yang2024pris}, 
  or diffusion models~\cite{yu2023cross}.
  Due to the significant payload, several conceptual differences exist between the two paradigms.
  Since the goal is to transfer semantic information, the Payload does not need to be perfectly encoded. Some errors are acceptable, 
  as long as the overall content of the secret image can be perceived.
  To do so, Baluja~\cite{baluja2019hiding} argues that ``{\it given the amount of information that we hide, we do not attempt to conceal the existence of a hidden message. 
  We do, however, [\dots] obfuscate the content of the hidden message.}'' 
  However, later papers presented schemes that aim at being undetectable to proper steganalysis methods, such as the HiNet model proposed by Jing et al.~\cite{jing2021hinet}, 
  which is presented as achieving ``{\it image hiding with large capacity and high security}''. 

  Despite the interest of such schemes, their security has not yet been thoroughly studied.
  In particular, to the best of our knowledge, Peng et al~\cite{peng2024keyless} is the only study
  that aims at detecting such Stego images. It does so in a supervised-learning setup, where they aim to train a surrogate model to extract the Payload image\footnote{Additionally, let us note that most of the papers doing image-in-image steganography include some detection results, using classical steganalysis methods -- see for instance~\cite{jing2021hinet,lu2021large}, where the schemes are shown to be very difficult to detect.}.
  In this paper, we propose to assess the security of image-in-image steganography schemes using 
  a non-supervised feature extraction method, as an input to a supervised classifier.
  This assessment first uses a novel dedicated steganalysis method that relies on independent component analysis (ICA) and principal component analysis (PCA) to perform blind source separation of the mixed signals. 
  In a second step, we also evaluate the detectability of these Stego images using classical steganalysis methods and show that they are highly vulnerable to such approaches.

  The rest of the manuscript is organized as follows.
  Section~\ref{sec:related_work} presents prior works related to image-in-image steganography and steganalysis.
  Section~\ref{sec:mixing} analyzes the mixing process induced by INN-based steganography schemes.
  Section~\ref{sec:methodology} details the proposed steganalysis method, relying on independent component analysis.
  Section~\ref{sec:experiments} reports experimental results and analysis.
  Finally, Section~\ref{sec:conclusion} concludes the paper.


\section{Prior works}
\label{sec:related_work}

  In this section, we detail the necessary background related to image-in-image steganography,
  and in particular to INN-based schemes.
  We also discuss the limitations of the state of the art in this field, which motivate the present work.

  Note that early proposals of image-in-image hiding consisted of jointly trained convolutional neural networks 
~\cite{baluja2017hiding,baluja2019hiding},
  where one neural network is trained to hide the Payload in the Cover, and another to recover it from the Stego.
  Similarly, Wu et al.~\cite{wu2018stegnet} base their embedding-decoding architecture on residual connections on 
  convolutional layers.

  \subsection{Invertible neural networks}

    \begin{figure}
      \centering
      \begin{tikzpicture}
        
        \node at (0,.5) (x1) {$\mathbf{x}^{l}_{1}$};
        \node at (0,-2) (x2) {$\mathbf{x}^{l}_{2}$};

        \node[draw=blue, thick, circle, fill=blue!25!white] at (2,-2) (plus1) {$+$};
        \node[draw=blue, thick, circle, fill=blue!25!white] at (3,.5) (dot) {$\odot$};
        \node[draw=blue, thick, circle, fill=blue!25!white] at (5,.5) (plus2) {$+$};

        \node[draw=violet, thick, fill=violet!50!white, minimum width=0.5cm, minimum height=0.5cm] at (2, -1) (phi) {$\phi$};
        \node[draw=violet, thick, fill=violet!50!white, minimum width=0.5cm, minimum height=0.5cm] at (3, -1.25) (rho) {$\rho$};
        \node[draw=violet, thick, fill=violet!50!white, minimum width=0.5cm, minimum height=0.5cm] at (5, -1) (eta) {$\eta$};

        \node[draw=red, thick, fill=red!50!white, minimum width=0.625cm, minimum height=0.375cm, inner sep=1pt] at (3, -0.375) (exp) {\scriptsize $\exp(\sigma(.))$};

        \node at (7,.5) (y1) {$\mathbf{x}^{l+1}_{1}$};
        \node at (7,-2) (y2) {$\mathbf{x}^{l+1}_{2}$};

        \draw[-latex, thick] (x1) -- (dot);
        \draw[-latex, thick] (x1) -| (phi);
        \draw[-latex, thick] (x2) -- (plus1);
        \draw[-latex, thick] (plus1) -| (rho);
        \draw[-latex, thick] (phi) -- (plus1);
        \draw[-latex, thick] (rho) -- (exp);
        \draw[-latex, thick] (exp) -- (dot);
        \draw[-latex, thick] (dot) -- (plus2);
        \draw[-latex, thick] (plus1) -- (y2);
        \draw[-latex, thick] (plus1) -| (eta);
        \draw[-latex, thick] (plus2) -- (y1);
        \draw[-latex, thick] (eta) -- (plus2);

      \end{tikzpicture}
      \caption{Illustration of the general architecture of the coupling layer of the INN in the HiNet model~\cite{jing2021hinet}. The complete INN is formed of 16 blocks. 
              The arrows indicate the direction of the forward pass.}
      \label{fig:inn_block}
    \end{figure}
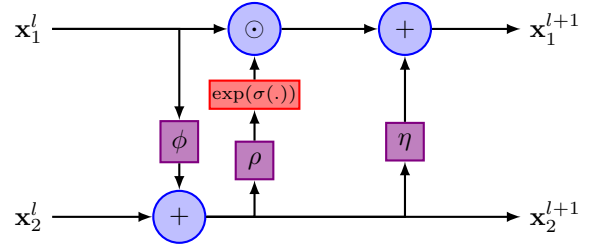

    Invertible Neural Networks are a class of neural networks designed to be bijective mappings between input and output spaces.
    They were first introduced by Dinh et al.~\cite{dinh2014nice}.
    Formally, an INN can be represented as a function $f: \mathbb{R}^n \rightarrow \mathbb{R}^n$, such that for every input $\mathbf{x} \in \mathbb{R}^n$, 
    there exists a unique output $\mathbf{y} = f(\mathbf{x})$ and vice versa, i.e., $\mathbf{x} = f^{-1}(\mathbf{y})$.
    This property of invertibility is achieved via the design of \emph{coupling layers}, 
    which split the input vector $\mathbf{x}$ into two parts $\mathbf{x}_1$ and $\mathbf{x}_2$, 
    and apply a series of transformations on them to produce two outputs. 
    As illustrated in Fig.~\ref{fig:inn_block}, 
    at coupling layer $l$, the outputs can be described as:
    \begin{equation}
      \begin{cases}
        \mathbf{x}^{l+1}_{1} = \mathbf{x}^{l}_{1} \odot \exp\left(\sigma\left(\rho(\mathbf{x}^{l+1}_{2})\right)\right) + \eta(\mathbf{x}^{l+1}_{2}), \\[6pt]
        \mathbf{x}^{l+1}_{2} = \mathbf{x}^{l}_{2} + \phi(\mathbf{x}^{l}_{1}),
      \end{cases}
    \end{equation}
    where $\phi$, $\rho$, and $\eta$ are learnable functions, often implemented as small neural networks, 
    $\sigma$ is the sigmoid activation function,
    and $\odot$ denotes the element-wise multiplication.
    Critically, the input can be computed simply by applying the opposite operations between the considered functions:
    \begin{equation}
      \begin{cases}
        \mathbf{x}^{l}_{1} = \left(\mathbf{x}^{l+1}_{1} - \eta(\mathbf{x}^{l+1}_{2})\right) \odot \exp\left(-\sigma\left(\rho(\mathbf{x}^{l+1}_{2})\right)\right), \\[6pt]
        \mathbf{x}^{l}_{2} = \mathbf{x}^{l+1}_{2} - \phi(\mathbf{x}^{l}_{1}).
      \end{cases}
    \end{equation}
    As a result, INNs are useful for applications requiring reversible transformations,
    such as image-in-image steganography, where the goal is to embed a secret image into a Cover image and to later recover it without loss.

    Rather than working on the color channels directly, the literature proposes to use a Haar wavelet decomposition of the 
    images~\cite{jing2021hinet,guan2022deepmih,yang2024pris}. Arguably, this helps the learning model to separate essential low frequencies and high-frequency information 
    that plays a lesser role in the overall semantics, and thus decreases the information loss.
    As we will show in section~\ref{sec:methodology}, we use this decomposition as a basis of our independent component analysis. 

  \subsection{Security related limitations}

\begin{figure}
      \centering
      \includegraphics[width=\columnwidth]{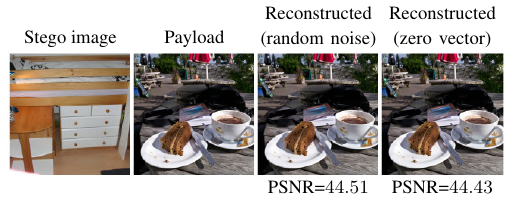}
      \vspace{-0.75cm}
      \caption{Secret image from a HiNet Stego image using different noise inputs.
               From left to right: Stego image, Payload image, revealed Payload with random noise, revealed Payload with zero noise.
               The PSNR  (in dB) is computed between the revealed secret images and the original secret image.}
      \label{fig:inn_reveal}
      \vspace{-0.25cm}
    \end{figure}

    Despite the potential of INN-based image-in-image steganography schemes,
    their security remains largely unexamined.
    In virtue of the Kerckhoffs' principle~\cite{shannon1949communication},
    a steganographic scheme should remain secure even if all its details, except for the secret key, are known to the adversary.
    However, most existing works on INN-based image-in-image steganography~\cite{jing2021hinet, lu2021large, xu2022robust} do not incorporate a secret key into their embedding and extraction processes.
    This omission raises significant security concerns, as it implies that an adversary with knowledge of the embedding algorithm could extract the hidden image without any secret information.
    Taking the example of the HiNet model~\cite{jing2021hinet},
    the architecture and weights of the INN entirely determine the embedding and extraction processes,
    and the choice of the input residual noise when revealing the secret image is seemingly irrelevant, as setting this noise to zero 
    already allows for a good reconstruction of the secret image, as shown in Fig.~\ref{fig:inn_reveal}.

    Furthermore, as highlighted by Lu et al.~\cite{lu2021large}: ``{\it in INNs, the basic invertible coupling layer is the additive affine transformations proposed by NICE}'' (Non-linear independent components estimation)~\cite{dinh2014nice}.
    This observation suggests that using the model as the key is not sufficient to ensure security, 
    as the coupling layers are based on well-known mathematical transformations that can be analyzed and potentially inverted by an adversary.
    In any case, as we will highlight in the following sections, even without leveraging nonlinear models such as NICE, we can extract meaningful components using linear ICA algorithms.

\section{Analyzing the mixing process of INN-based schemes}
\label{sec:mixing}

  In this section, we present a preliminary analysis of the mixing process performed by INN-based images-in-image
  steganography. We focus on the HiNet model~\cite{jing2021hinet},
  which is one of the most representative and widely accepted in the literature.
  Our goal is to understand the nature of the mixing process performed by the INN, 
  and to assess what information from the Payload image is carried by the Stego image.

  To do so, we conduct a simple analysis. We first apply a discrete wavelet transform (DWT) to the Cover, Payload, and Stego images.
  We can then visualize the difference between the DWT sub-bands of the Cover and the Stego images.
  This difference amounts to the embedding changes brought by the INN, and thus gives us insights
  into the information from the Payload added to the Cover image to produce the Stego one.

\begin{figure}
    \centering
    \includegraphics[width=0.9\columnwidth]{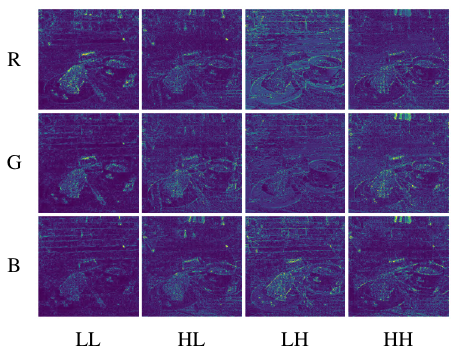}
    \vspace{-0.25cm}
    \caption{Embedding changes in each sub-band of the DWT of the stego image, computed as the difference with the DWT sub-bands of the cover.}
    \label{fig:dwt_diff}
    \vspace{-0.25cm}
  \end{figure}

  This difference is illustrated in Fig.~\ref{fig:dwt_diff}, using the Cover, Payload, and Stego images shown in Fig.~\ref{fig:method_overview}.
  Critically, we see that the embedding changes are not random noise, but rather carry significant information from the Payload image.

\begin{figure}
    \centering
    \includegraphics[width=0.9\linewidth]{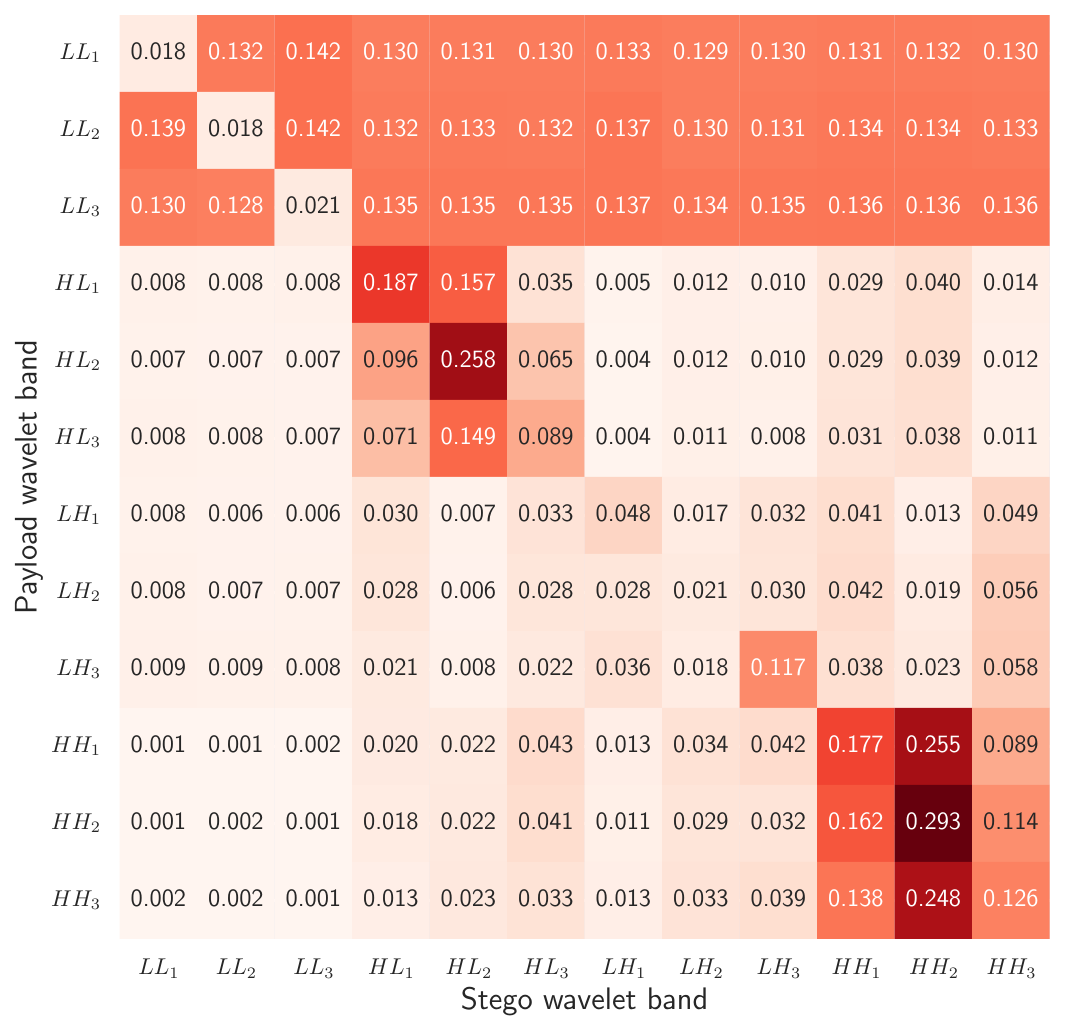}
    \vspace{-0.25cm}
    \caption{Correlation matrix between the sub-bands of the DWT of the payload image and the difference between the
    ones of the Cover and the Stego images, when using the HiNet model~\cite{jing2021hinet}. The Pearson correlation coefficient is used. 
    It shows that the low-frequency sub-bands of the payload are mostly embedded in the Stego image.}
    \label{fig:dwt_corr}
    \vspace{-0.25cm}
  \end{figure}

  To assess how much, on average, the embedding changes carry information from the Payload, we 
  propose to compute the correlation of each sub-band of the DWT of the Payload 
  with the embedding changes of each sub-band of the DWT of the Stego image.
  The results of this analysis are illustrated in Fig.~\ref{fig:dwt_corr}, averaging results over 100 Cover-Payload-Stego triplets.
  We can see that the embedding changes are highly correlated with the low-frequency components of the Payload image,
  and overall carry less information from the high-frequency components of the Payload image. As a side-note, the $HL$ and $HH$ sub-bands of the Payload are respectively correlated to the $HL$ and $HH$ sub-bands of the Cover image.

\vspace{-0.25cm}
\section{Proposed methodology}
\label{sec:methodology}

    In this section, we detail the proposed steganalysis method.
    We begin by recalling the main principles of independent component analysis, which underpin our method.
    We then motivate using PCA as a preprocessing step to select the most relevant sub-bands of the wavelet decomposition for ICA.
    We then detail the proposed features for training a simple classifier.
    
  \subsection{Independent Component Analysis}

    Independent Component Analysis (ICA) is a computational technique for separating a multivariate signal into additive, independent non-Gaussian components.
    It is beneficial in blind source separation problems, where the goal is to recover source signals from observed mixtures without prior knowledge of the mixing process.
    The fundamental assumption of ICA is that the source signals are statistically independent and non-Gaussian.

    Mathematically, ICA~\cite{jutten1991blind} can be described as follows.
    Let $\mathbf{x} = [x_1, x_2, \ldots, x_m]^T$ be the observed random vector, 
    which is assumed to be a linear mixture of $n$ independent source signals $\mathbf{s} = [s_1, s_2, \ldots, s_n]^T$: $\mathbf{x} = A \mathbf{s}$,
    where $A$ is an unknown mixing matrix.
    The objective of ICA is to estimate both the mixing matrix $A$ and the source signals $\mathbf{s}$ from the observed data $\mathbf{x}$.
    Various algorithms exist for performing ICA~\cite{linsker1988self, cardoso1993blind, bach2002kernel}. 
    In this work, we rely on the FastICA algorithm proposed by Hyvärinen~\cite{hyvarinen2000independent}.
    It consists in an iterative fixed-point method that maximizes the non-Gaussianity of the estimated components.
    We refer to~\cite{hyvarinen2000independent}[Section 6.1] for more details. 
    We use the FastICA implementation proposed in the \texttt{scikit-learn} library with the default parameters.

  \vspace{-0.25cm}
  \subsection{Applying the ICA to Stego image detection}

    \begin{figure}
      \centering
      \includegraphics[width=0.95\columnwidth]{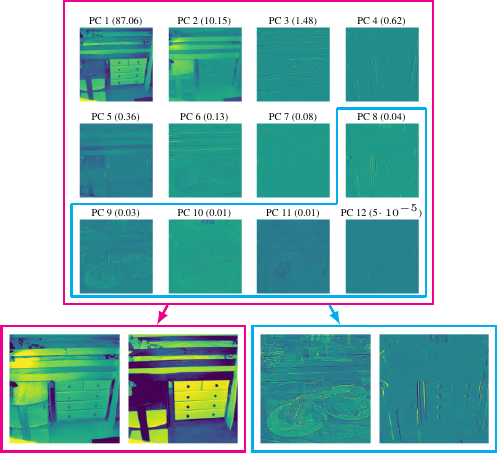}
      \caption{Comparison of the ICA sub-bands obtained when using all the PCA components, 
              or only the 5 smallest ones. 
              The variance explained by each component is given (in \%).} 
      \label{fig:pca_components}
      \vspace{-0.25cm}
    \end{figure}

    To work, the ICA requires that we observe at least as many mixtures as sources.
    However, in the most realistic setup where we do not have access to the steganographic model, 
    we can only assume that we have access to a single Stego image, which is a single mixture of the 
    two signals we are trying to extract: the Cover image and the Payload image.

    To address this issue, we can leverage insights from the analysis in Section~\ref{sec:mixing}.
    Indeed, we have seen that the embedding changes brought by the INN-based steganographic scheme
    consisted in adding different information from the sub-bands of the Payload into the sub-bands of the Cover (see Fig.~\ref{fig:dwt_corr}).
    Since the mixtures are different between the different sub-bands of the Stego image, we can consider them as the different mixed signals
    as required by the ICA.

    Furthermore, we can further refine this approach by selecting only specific sub-bands for ICA.
    To extract the most relevant sub-bands, we propose to perform a principal component analysis (PCA) on the 12 wavelet sub-bands 
    (4 sub-bands per color channel) obtained from the DWT of the images.
    The rationale is that the most significant components of the PCA capture the image's dominant structures, i.e., the content of the Cover.
    On the other hand, the less significant components are more likely to contain the modifications introduced by the embedding process, 
    i.e., the secret image.
    This is illustrated in Fig.~\ref{fig:pca_components}, where we can see that the small components of the PCA clearly contain 
    information related to the Payload. As a result, when performing the ICA, we can extract visually more meaningful components
    when using only the small PCA components rather than all of them.

  \vspace{-0.25cm}

  \subsection{Detection scheme}

    Once the ICA has been performed on the selected PCA components of the wavelet decomposition of the Stego image, we obtain a pair of independent components, which are shown to contain different semantic information in Fig.~\ref{fig:pca_components}.
    On these components, which we denote $\mathbf{c}_1$ and $\mathbf{c}_2$, we then compute and concatenate simple statistical features, namely the first four moments of the distributions of their coefficients:
    \begin{equation}
      \text{Feature} = \left\{ \mu_{1}, \mu_{2}, \sigma_{1}, \sigma_{2}, \gamma_{1}, \gamma_{2}, \kappa_{1}, \kappa_{2} \right\},
    \end{equation}
    where $\mu_{i}$, $\sigma_{i}$, $\gamma_{i}$ and $\kappa_{i}$ denote the mean, standard deviation, skewness and kurtosis of the coefficients of component $i$, respectively.
    Finally, a Gaussian SVM is trained on these features to perform Stego/Cover classification.

    Fig.~\ref{fig:method_overview} summarizes the overall scheme.
    It starts with the DWT of the input image, 
    followed by a PCA to select the most relevant sub-bands.
    The ICA is then performed on these sub-bands to extract two independent components.
    Finally, the first four moments of the  coefficient distributions for these components are computed and used as input features for an SVM classifier.
    We illustrate in this figure the different steps with one Cover and one Stego example, 
    to show the differences in the extracted components and features.


\section{Experiments and results analysis}
\label{sec:experiments}

  \subsection{Experimental setup}

    We propose the following experimental setup to evaluate our method.
    For the steganographic schemes, we consider five different models:
    Hinet~\cite{jing2021hinet}, PRIS~\cite{yang2024pris} and DeepMIH~\cite{guan2022deepmih} as INN-based schemes,
    as well as Baluja~\cite{baluja2019hiding} and Weng et al.~\cite{weng2019high}.
    Note that we use the publicly available weights for all models except for PRIS, which we trained using the official 
    repositories\footnote{\href{https://github.com/yanghangAI/PRIS}{github.com/yanghangAI/PRIS}}.
    For all the models, we generate 2500 Stego images using the COCO dataset images~\cite{lin2014microsoft} 
    used by Guillaro et al.~\cite{guillaro2025bias}, which also forms our base for Cover images. All images are $512 \times 512$ color images.

    To know which PCA components to select, we perform a grid search on each pair of PCA components, 
    and choose the pair that empirically leads to the best detection performance for the HiNet images.
    Using components 9 and 11 lead to the best results, which are thus used in the following experiments.

    All classifiers are trained using 
    5-fold cross-validation to ensure robust results, with a balanced number of Cover and Stego images.

  \vspace{-0.25cm}
  \subsection{Results of the proposed method}

    We now evaluate the detection performance of the proposed scheme.
    The results are reported in Table~\ref{tab:detection_results}.
    We can see that the proposed method can achieve good detection accuracy, in particular on the 
    INN-based schemes, with up to $84.6\%$ accuracy on PRIS.
    However, we can also see that the non-INN schemes are less detectable. 
    This can be explained by the fact that these methods process the pixels directly rather than the DWT coefficients.

    Considering how synthetic and interpretable the proposed method is, these results are quite interesting,
    and highlight the high detectability of such hiding schemes. 

    \begin{table}
      \centering
      \small
      \begin{tabular}{|c||c|c|c|c|c|}
        \hline
        Method & HiNet & PRIS & DeepMIH &  Baluja & Weng \\ 
        \hline\hline
        Acc (\%) & $80.31$ & $84.62$ & $82.58$ & $61.83$ & $74.96$ \\ 
        \hline
        Std ($\pm$\%) & $0.76$ & $0.56$ & $1.03$ & $0.74$ & $1.60$ \\ 
        \hline
      \end{tabular}
      \caption{Detection accuracy and standard deviation (in \%) of the proposed scheme on different image-in-image steganography schemes.}
      \label{tab:detection_results}
      \vspace{-0.25cm}
    \end{table}


  \vspace{-0.25cm}
  \subsection{Robustness to classical steganalysis}

    We now evaluate the detectability of the considered steganographic models 
    using classical steganalysis features, namely the Spatial Rich Model (SRM)~\cite{fridrich2012rich} 
    combined with a support vector machine (SVM) classifier. 
    This method is expected to trade off some of the interpretability of the input features for 
    a higher detection accuracy.

    The results are reported in Table~\ref{tab:classical_steganalysis}.
    We can see that all the considered steganographic schemes are highly detectable using this method,
    with detection accuracy above $99\%$ for all methods, except for Baluja's method, which is still detected with $80\%$ accuracy.
    This further highlights the high vulnerability of such image-in-image steganography schemes. 

    \begin{table}
      \centering
      \small
      \begin{tabular}{|c||c|c|c|c|c|}

        \hline
        Method & HiNet & PRIS & DeepMIH &  Baluja & Weng \\ 
        \hline\hline
        Acc (\%) & $99.02$ & $99.96$ & $99.92$ & $80.06$ & $99.64$ \\
        \hline
        Std ($\pm$\%) & $0.29$ & $0.05$ & $0.07$ & $0.85$ & $0.15$ \\
        \hline
      \end{tabular}
      \caption{Detection accuracy and standard deviation (in \%) of classical steganalysis (SRM + SVM) on different image-in-image steganography schemes.}
      \label{tab:classical_steganalysis}
      \vspace{-0.25cm}
    \end{table}


\vspace{-0.25cm}
\section{Conclusion and perspectives}
\label{sec:conclusion}

  In this study, we investigated the vulnerability of image-in-image steganography schemes,
  particularly those based on invertible neural networks (INNs).
  We proposed a novel steganalysis method based on the first four moments of the distribution of coefficients of the 
  ICA of the DWT of the images after PCA.
  This very synthetic and highly interpretable method was shown to achieve interesting detection performance, 
  highlighting the high detectability of such Stego images.
  Furthermore, we also showed that classical steganalysis could achieve high detection rates on such Stego images,
  further emphasizing their vulnerability.

  We hope that this study will motivate the design of more secure image-in-image steganography schemes in the future.
  For example, introducing a secret key in the hiding scheme, combined with a loss measuring detectability
  would be a first step in this direction.

    %

\vfill\pagebreak

\section{Acknowledgements}

This work was supported by the French
government grant managed by the Agence Nationale de la
Recherche under the France 2030 program, reference ANR22-PECY-0011.


\bibliographystyle{IEEEbib}
\bibliography{refs}

\end{document}